**Optimal population-level infection detection strategies for malaria control and elimination in a spatial model of malaria transmission**

Short title: Identifying people with subpatent malaria infections

Jaline Gerardin[1]*, Caitlin A. Bever[1], Busiku Hamainza[2], John M. Miller[3], Philip A. Eckhoff[1], Edward A. Wenger[1]

[1]Institute for Disease Modeling, Bellevue, Washington, United States of America

[2]National Malaria Control Centre, Lusaka, Zambia

[3]PATH Malaria Control and Elimination Partnership in Africa (MACEPA), Lusaka, Zambia

*Corresponding author. Email: jgerardin@intven.com (JG)

# Abstract

Mass campaigns with antimalarial drugs are potentially a powerful tool for local elimination of malaria, yet current diagnostic technologies are insufficiently sensitive to identify all individuals who harbor infections. At the same time, overtreatment of uninfected individuals increases the risk of accelerating emergence of drug resistance and losing community acceptance. Local heterogeneity in transmission intensity may allow campaign strategies that respond to index cases to successfully target subpatent infections while simultaneously limiting overtreatment. While selective targeting of hotspots of transmission has been proposed as a strategy for malaria control, such targeting has not been tested in the context of malaria elimination. Using household locations, demographics, and prevalence data from a survey of four health facility catchment areas in southern Zambia and an agent-based model of malaria transmission and immunity acquisition, a transmission intensity was fit to each household based on neighborhood age-dependent malaria prevalence. A set of individual infection trajectories was constructed for every household in each catchment area, accounting for heterogeneous exposure and immunity. Various campaign strategies—mass drug administration, mass screen and treat, focal mass drug administration, snowball reactive case detection, pooled sampling, and a hypothetical serological diagnostic—were simulated and evaluated for performance at finding infections, minimizing overtreatment, reducing clinical case counts, and interrupting transmission. For malaria control, presumptive treatment leads to substantial overtreatment without additional morbidity reduction under all but the highest transmission conditions. Compared with untargeted approaches, selective targeting of hotspots with drug campaigns is an ineffective tool for elimination due to limited sensitivity of available field diagnostics. Serological diagnosis is potentially an effective tool for malaria elimination but requires higher coverage to achieve similar results to mass distribution of presumptive treatment.

## Author summary

Millions of people worldwide live at risk for malaria, a parasitic infectious disease transmitted by mosquitoes. Great progress has been made in reducing malaria burden in recent years, and many regions are now devising strategies for elimination. One way to eliminate malaria is to deplete the reservoir of parasites in human hosts by treating large groups of people with antimalarial drugs. However, current field diagnostics are not sensitive enough to correctly identify all infected individuals. Presumptively administering antimalarial drugs to whole populations will effectively clear infections but can also lead to substantial overtreatment and encourage the evolution of drug resistance in parasites. We might be able to predict which individuals who test negative are actually infected based on whether their household members and neighbors are testing positive. Using a mathematical model of malaria immunity acquisition and a spatial dataset of malaria prevalence in southern Zambia, we simulate strategies of identifying infected individuals and compare each strategy's ability to deplete the infectious reservoir and avoid overtreatment. We make different recommendations for optimal strategies depending on a region's malaria prevalence.

## Introduction

Malaria is a widespread infectious disease caused by *Plasmodium* parasites and leads to over half a million deaths each year, mostly in children under five years of age [1]. As global burden has decreased dramatically over the past decade, local elimination of malaria is within sight for more and more endemic areas. Regional elimination of malaria requires interrupting transmission between humans and mosquito vectors, and understanding the requirements for elimination is crucial for avoiding costly operations that are unlikely to succeed [2,3].

Because the reservoir of malaria parasites lies in the human population, campaigns with antimalarial drugs can interrupt transmission under certain conditions [4]. Testing such campaigns in the field is resource-intensive, and computational models have been used to describe how factors such as campaign coverage, local malaria transmission intensity, and individual compliance with drug regimens affect campaign outcomes [5-9]. Mathematical modeling has shown that repeated annual campaigns of antimalarial drugs with high coverage can achieve local elimination in low- to moderate-transmission settings. Modeling has also confirmed that mass drug administration (MDA), where all individuals in a population are presumptively treated, can be substantially more effective than requiring positive diagnosis prior to treatment, as subpatent infections can constitute a substantial portion of the infectious reservoir [10,11].

Although drug campaigns can be effective, large-scale interventions with antimalarials pose several potential drawbacks. Dosing a large population will accelerate the emergence of drug-resistant parasites [12,13]. Parasite resistance to both artemisinin and partner drug in artemisinin-based combination therapies has been observed in Southeast Asia, and spread of resistance to Africa would be catastrophic [14]. Repeated rounds of campaigns can lead to community fatigue and widespread unnecessary suffering of drug side effects, and high community coverage has been shown to be vital to campaign success [4,5]. Lastly, treating people who are uninfected and not at risk for infection is a waste of valuable resources.

Malaria transmission can be highly heterogeneous between neighboring villages and within the same village [15]. Selective targeting of hotspots of transmission has been predicted to improve results of vector control interventions [16]. However, it remains unknown whether current field diagnostics are adequate tools for defining hotspots, whether targeting hotspots with drug campaigns can achieve elimination, and how outcomes of targeted approaches compare with non-targeted approaches such as MDA.

Individuals living in malaria-endemic areas develop partial immunity to malaria, leading to asymptomatic, low-density infections that are difficult to detect but continue to infect mosquitoes [17]. Mass screen-and-treat (MSAT) campaigns, where only individuals who test positive are treated with antimalarial drugs, have historically failed to achieve elimination due to the limited sensitivity of rapid diagnostic tests (RDTs) currently used in the field [18-21]. In an ideal scenario, a cheap, fast, and sensitive field diagnostic would increase effectiveness of MSAT campaigns to near parity with MDA campaigns while avoiding overtreatment of uninfected individuals [22-24]. In the absence of such a diagnostic, smart campaigns should be designed to treat as many subpatent infections as possible while simultaneously treating few uninfected individuals [25].

Observed patterns in spatial heterogeneity in infection status may allow campaigns to effectively target subpatent infections based on proximity to an index case. Since members of the same household and close neighbors likely experience similar exposure [21,26-28], conducting a focal MDA (fMDA) around a confirmed-positive case may be a sound strategy for detecting subpatent infections in the face of limited sensitivity of RDTs [29,30]. It remains unknown whether fMDAs can approach MDA campaigns' effectiveness at interrupting transmission and how the size of the fMDA treatment area should be selected. The amount of overtreatment that can be averted by conducting fMDA campaigns rather than MDA campaigns is also unknown. Predicting both of these effects requires coupling spatial knowledge of regional heterogeneity in malaria exposure with a validated model of immunity acquisition in humans and transmission between humans and vectors.

Here we present for the first time modeling of malaria transmission on an operationally relevant scale accounting for household-scale levels of heterogeneity in transmission intensity and immunity. We use a previously described model of malaria transmission, including a within-host model of immunity calibrated to age-stratified prevalence, incidence, and parasite density data from endemic settings, to estimate household exposure based on a spatial dataset of individual infection status from Southern

Province, Zambia, where operations teams are currently carrying out mass drug campaigns [19]. Campaigns with antimalarial drugs are simulated employing a variety of infection detection strategies, including MDA, MSAT, fMDA, reactive case detection (RCD), pooled polymerase chain reaction (PCR), and a hypothetical serological diagnostic. We compare the strategies' ability to avert clinical cases and interrupt transmission with minimal overtreatment of uninfected individuals.

## Results

**Spatial clustering of RDT positivity**

A detailed survey of RDT prevalence by age and household was conducted in Southern Province, Zambia, in June-July 2012 at the end of the transmission season (Fig 1A) [19]. Four representative health facility catchment areas (HFCAs) experiencing a wide range of malaria transmission intensity were selected for this analysis. Overall RDT prevalence in the four HFCAs spanned 1.4 to 49%, but varied widely between households in the same HFCA, particularly in the Bbondo and Munyumbwe HFCAs where prevalence was intermediate (Fig 1B).

**Fig 1. RDT-positive infections are clustered within four HFCAs in Southern Province, Zambia.** (A) Study area with June-July 2012 RDT prevalence. (B) Household RDT positive rate varied spatially within a HFCA. (C) RDT positive individuals within a HFCA were clustered within households and within 50m. Conditional probabilities of being RDT+ were calculated as the fraction of an RDT+ individual's family members who were also RDT+, the fraction of people who were RDT+ within a 50m radius but not within the household of an RDT+ individual, and the fraction of people who were RDT+ between 50m

and 200m of an RDT+ individual. Bars indicate 95% confidence intervals. (D) RDT positive rate varied with age within each HFCA. Shaded areas indicate 95% confidence intervals.

In all four HFCAs, we observed clustering of RDT positive cases within households—individuals were more likely to be RDT positive if someone else in their household was RDT positive (Fig 1C). Individuals were also more likely to be RDT positive if someone within 50m, but not in their household, was RDT positive. The clustering of RDT positivity within 50m held even in the Gwembe HFCA, suggesting that a small amount of endemic transmission persists in Gwembe and not all infections are imported.

To predict the outcome of various infection detection strategies on reducing transmission, we constructed a set of synthetic households for each HFCA made up of simulated individuals that reflected the geography and demographics of RDT positivity observed in the reference data for the HFCA (Fig 1, Table 1, Fig S1). We assumed that members of each household experienced the same transmission intensity, but households within an HFCA could experience a different transmission intensity. Transmission intensity of each household was determined by comparing the RDT prevalence by age of the household's neighborhood to reference curves from simulations of known transmission intensity (see Methods, Fig S2).

**Table 1. HFCA demographics used in simulation**

| HFCA | Households | Population | RDT prevalence | Fraction under 15 years |
| --- | --- | --- | --- | --- |
| Gwembe | 613 | 2084 | 1.4% | 48% |
| Bbondo | 296 | 1462 | 9.2% | 54% |
| Munyumbwe | 1507 | 7209 | 27% | 56% |
| Sinamalima | 1599 | 9900 | 49% | 55% |

In Gwembe and Sinamalima HFCAs, nearly all households experienced very low or very high transmission respectively (Fig 2A). Households in Bbondo and especially Munyumbwe HFCAs were more heterogeneous, and spatial patterns of high and low transmission intensity mirrored the household RDT positive rates (Fig 2B). The spatial clustering of RDT positivity within and near households suggested that fMDAs may be a good strategy for infection detection.

**Fig 2. Fitting of transmission intensities to individual households in four HFCAs shows a prominent role for transmission by subpatent individuals.** (A) Distribution of fitted household transmission intensities by HFCA. Shaded areas indicate 95% confidence intervals from 100 stochastic realizations. (B) Household transmission intensity varied spatially within an HFCA. Geometric mean transmission intensity observed over 100 stochastic realizations. (C) Asexual parasite prevalence and infectious potential of constructed populations by HFCA, age, and detectability of asexual parasites by current RDTs, improved RDTs, and PCR, normalized to population 1000, on June 15. Here the infectious potential was defined as the number of mosquitoes that would be infected if 1000 mosquitoes were to feed on a village of 1000 people. Results shown are means of 100 stochastic realizations.

**The infectious reservoir of simulated households constructed based on RDT prevalence data**

Since the asexual parasite density and infectiousness of each simulated individual was known, the true parasite prevalence and infectious potential of each HFCA could be estimated from the simulated households (Fig 2C). Infectious potential, a proxy for the infectious reservoir of malaria parasites in a human population, was defined as the number of mosquitoes that would be infected if 1000 mosquitoes were to feed on a village of 1000 people and accounts for heterogeneity in individual infectiousness due to parasite density and preference for mosquitoes to bite larger people.

We found that all four HFCAs had substantial rates of subpatent infection. Low-density infections were four times as common as RDT-detectable infections in the Gwembe HFCA, where prevalence was the lowest, consistent with previous observations of low-density infections in low-prevalence seasonal settings [31,32]. Under low-transmission conditions in our model, infections observed during June-July were three to six months old and past peak parasite levels. In contrast, June-July infections under high transmission were more likely to be newer and composed of multiple infections, leading to higher parasite density. Higher density infections acquired in low-transmission settings during the rainy season were also more likely to have triggered symptoms and hence treatment due to weaker host immunity.

Although low-density infections are less infectious than high-density infections, these subpatent infections were comprised a substantial portion of the infectious potential in all four HFCAs. Targeting infections with an RDT-based MSAT would thus be highly unlikely to lead to elimination at any level of transmission intensity. Even improvement of RDT sensitivity by an order of magnitude from 100 parasites per µL to 10 parasites per µL would still leave a nontrivial amount of remaining infectious potential after an MSAT campaign.

**Success of infection detection strategies at reducing onward transmission**

Depleting the infectious reservoir was highly dependent on coverage and somewhat dependent on infection-detection strategy (Fig 3A, Table 2). For all HFCAs, MDA was the most successful at depleting the infectious reservoir, MSAT the least successful, and other infection detection strategies fell in between MDA and MSAT. Since dry season infections were more likely to be low-density in low-transmission settings than in high-transmission settings, MSAT was comparatively least effective at depleting the infectious reservoir at low prevalence, achieving only 40% of MDA's effect in Gwembe HFCA, and most effective at high prevalence, achieving 70% of MDA's effect in Sinamalima HFCA.

**Fig 3. Infection detection strategies differ in ability to reduce transmission.** (A) Success of infection detection strategies at depleting the infectious reservoir. Fraction of infectious reservoir eliminated was defined as the decrease in infectious potential integrated over 30 days post-campaign. Results shown are means of 100 stochastic realizations per coverage level. (B) Success of infection detection strategies at averting new infections. Results shown are means of 100 stochastic realizations per coverage level. HFCA populations were normalized to 1000.

**Table 2. Infection detection strategies simulated**

| Test-independent strategies | | |
|---|---|---|
| MDA | Treat all individuals in HFCA | All HFCAs |
| **RDT-dependent strategies** | | |
| MSAT | Treat all RDT-positive in HFCA | All HFCAs |
| fMDA within household | Treat all individuals in same household of an RDT-positive | All HFCAs |
| fMDA within 50m | Treat all individuals within 50m of an RDT-positive | All HFCAs |
| fMDA within 200m | Treat all individuals within 200m of an RDT-positive | All HFCAs |
| **Fever- and RDT-dependent strategies** | | |
| Snowball RCD | Define a clinical case as temperature > 38.5°C. Treat all individuals within 200m of a clinical case. Also test with RDT all individuals within 200m of a clinical case, and treat all individuals within 200m of RDT-positives | All HFCAs |
| **PCR-dependent strategies** | | |
| Pooled PCR | Pool 20μL blood samples from 60-220 neighbors, test pooled samples with qPCR (0.75 parasites/μL sensitivity), treat all individuals in pools that test positive (**Fig S3**) | Gwembe, Bbondo |
| **Serology-dependent strategies** | | |
| MSAT with serological diagnostic | Treat all individuals who have experienced infection in the previous 12 months | Gwembe, Bbondo |
| fMDA within household with serological diagnostic | Treat all individuals in the same household as someone who has experienced infection in the previous 12 months | Gwembe, Bbondo |

To evaluate the reduction in transmission after deploying drug campaigns, we estimated the expected number of new infections that would be seeded in humans due to vectors becoming infected in the first 30 days post-campaign (Fig 3B). We assumed that vectors tended to bite in the same neighborhood and that individuals who had received treatment during the campaign were protected from reinfection (see Methods). Compared with outcomes from a non-prophylactic drug (Fig S4A), campaigns with a long-lasting prophylactic averted more new infections at moderate coverage, especially for MDA and other scenarios where a large fraction of the population was treated.

RDT-positive infections were more infectious than subpatent infections and, during the dry season, more likely to occur in households with a history of higher exposure. MSAT campaigns and other RDT-dependent infection detection strategies were therefore more effective at averting new infections than might be predicted from their effectiveness at depleting the infectious reservoir relative to MDA campaigns. At higher prevalence and higher coverage, fMDA strategies were just as effective as MDA at reducing onward transmission. An order of magnitude improvement of RDT sensitivity from 100 parasites per µL to 10 parasites per µL was insufficient for increasing the efficacy of RDT-dependent infection detection strategies up to levels seen with MDA in low-prevalence areas (Fig S5).

**Most effective infection detection strategies for malaria control**

In a control scenario where drug campaigns aim to reduce clinical incidence, we imagined that overtreatment was especially to be avoided, particularly in low-transmission settings, as it confers little benefit and may accelerate the rate of parasite resistance to antimalarial drugs. Fig 4A shows receiver operating characteristic (ROC) curves of fraction of infected individuals treated (true positive rate) vs fraction of uninfected individuals treated (false positive rate) for each of the infection detection strategies.

**Fig 4. Infection detection strategies differ in ability to minimize overtreatment.** (A) Success of infection detection strategies at finding infected individuals while avoiding overtreatment of uninfected individuals. Results shown are means of 100 stochastic realizations per coverage level. Ticks indicate 20% steps in coverage. (B) Success of infection detection strategies at averting clinical cases while minimizing overtreatment at 80% coverage. Results shown are means and 95% confidence intervals of 100 stochastic realizations per coverage level. HFCA populations were normalized to 1000.

MDA was agnostic to individual infection status, treating infected and uninfected individuals at the same rate as coverage increases. MSAT could not treat uninfected individuals, and limited sensitivity of the RDT diagnostic resulted in at most 50% of infected individuals receiving treatment with an MSAT campaign; MSAT found the highest fraction of infected individuals in Sinamalima HFCA where prevalence was high. The remaining infection detection strategies, the fMDAs and snowball RCD, fell between MSAT and MDA and in some cases exhibited favorable ROC curves, indicating a high rate of treating positives while minimizing treating negatives. For fMDAs, ROC curves decreased in favorability with increasing HFCA prevalence.

Focal MDAs were successful at avoiding overtreatment in all but the highest-prevalence HFCAs. Within-household fMDA and within-50m fMDA showed similar behavior, as households were sparse at 50m (Fig S1B), while expanding the treatment radius to 200m resulted in much more overtreatment without capturing nearly as many additional infections. Snowball reactive case detection resulted in more overtreatment per infection detected than within-200m fMDA in all four HFCAs.

To compare rates of clinical case prevention and overtreatment across HFCAs, we normalized populations to 1000 people and fixed coverage at 80%, a high but achievable rate (Fig 4B). Because transmission was so low in Gwembe HFCA, any mass campaign would avert only a handful of clinical cases: MSAT averted on average two clinical cases and MDA averted five, with the remaining infection detection strategies falling in between. Yet an MDA campaign would result in treating over 700 people who were uninfected, and those individuals derived little benefit from prophylactic effects given the low rate of transmission. Within-household fMDA, the infection-detection strategy that resulted in the least overtreatment next to MSAT, required overtreatment of nearly 50 individuals to avert less than one clinical case. These high rates of overtreatment suggested that MSAT might be the only reasonable option for mass treatment for malaria control in low-prevalence areas despite MSAT's relative inability to deplete the infectious reservoir.

In the Bbondo HFCA, within-household fMDA averted seven more clinical cases and overtreated 125 people than MSAT. Averting another five more clinical cases would require an MDA campaign that overtreated 475 more people.

In the Munyumbwe and Sinamalima HFCAs, within-household fMDA performed nearly as well as MDA at averting clinical cases, and for Munyumbwe, fMDA resulted in much lower numbers of overtreated people than MDA did. In Sinamalima, rates of overtreatment with fMDA were nearly comparable to those of MDA, and within-household fMDA resulted in 20 more cases averted than MSAT. In a high-prevalence site like Sinamalima, other factors such as costs or logistics would help decide whether fMDA or MDA is the best course of action.

**Most effective infection detection strategies for malaria elimination**

The Gwembe and Bbondo HFCAs were considered for elimination scenarios. In addition to the six infection detection strategies discussed above, we simulated pooled PCR, serological MSAT, and serological within-household fMDA to test strategies more appropriate for low-transmission regions. The serological tests were modeled as hypothetical diagnostics that report whether an individual has experienced infection at any point in the previous twelve months. We measured the probability of less than 1 new infection per 1000 people arising from vectors infected during the 30 days post-campaign as a proxy for elimination.

MDA was the most effective strategy for elimination, leading to high probability of less than 1 onward infection at lower coverage levels than the other infection detection strategies (Fig 5). However, pooled PCR and serological diagnostics could also be highly effective as long as coverage was high. MSAT with a serological diagnostic was especially promising as we predicted it to be efficient at avoiding overtreatment.

**Fig 5. MDA and sensitive serological diagnostics are the most effective detection strategies for malaria elimination.** (A) Probability that less than one new infection is seeded from vectors infected in the 30 days post-campaign. Results shown are means of 1000 stochastic realizations per coverage level. (B) Success of infection detection strategies at finding infected individuals while minimizing overtreatment. Results shown are means of 1000 stochastic realizations per coverage level. Ticks indicate 20% steps in coverage.

For pooled PCR, we grouped each HFCA into neighborhood pools consisting of 60-220 individuals per pool (Fig S3). Individuals contributed a dried blood spot to a pooled sample, and MDA within the pool was triggered if the pooled sample tested positive. In the Gwembe HFCA, pooled PCR led to lower overtreatment than MDA at the same level of coverage. However, even at 100% coverage, pooled PCR could not reliably find all infections due to the detection limit of pooled PCR. If a particular pool contained infected individuals but was not triggered for MDA, neighbors within the pool were vulnerable to onward transmission as no one in the pool received cure or prophylaxis. When we relaxed the assumption that vectors tend to transmit in households close to their site of infection, and instead allowed infected vectors to bite individuals anywhere within the HFCA, pooled PCR was able to achieve high probability of interrupting onward transmission, requiring higher population coverage than MDA but less coverage than serological-based fMDA (Fig S6).

In Bbondo HFCA, all PCR pools always contained enough parasites to trigger MDA within the pool. Pooled PCR became *de facto* MDA, indicating that performing pooled PCR would be a waste of resources as MDA is cheaper and easier to administer. Unless a region is extremely heterogeneous, where a subregion experiences no transmission at all while another experiences a moderate amount, and vectors cannot migrate between heterogeneous areas, we anticipate that pooled PCR is an inferior strategy to MDA. Neither of the lower-transmission HFCAs in this study showed such stark heterogeneity, but ongoing control efforts may push these regions into a regime where pooled PCR would be highly effective.

Of the RDT-based strategies, only within-200m fMDA showed any promise for elimination, and only with very small probability for the Bbondo HFCA at 100% coverage. As parasite densities were slightly higher in Bbondo than in the Gwembe HFCA due to higher levels of transmission, RDTs were more likely to identify infected individuals to seed the fMDA foci. However, fMDA at 200m did lead to substantial overtreatment compared to serology-dependent infection detection strategies. For all infection detection strategies, a long-lasting prophylactic improved the chances of no onward transmission (**Fig S4B**), and strategies such as pooled PCR that led to more overtreatment could outperform serological-based strategies at promoting elimination due to herd protection effects.

## Discussion

The selection of infection-detection strategy for a mass drug campaign depends on many factors, among them local transmission intensity, cost, operational feasibility, and population receptiveness. In this study, we compare the effectiveness of MDA, MSAT, fMDA, RCD, pooled PCR, and hypothetical serological diagnostics at averting clinical cases and reducing onward transmission with minimal overtreatment of uninfected individuals (Table 3).

**Table 3. Recommended infection detection strategies for malaria control and elimination**

|  | Control |  | Elimination |  |
|---|---|---|---|---|
| HFCA, prevalence | Recommend | Not recommend | Recommend | Not recommend |
| Gwembe, 1.4% | MSAT | MDA, snowball, all fMDAs | MDA, serology-based household fMDA if coverage > 80%, serology-based MSAT if coverage > 90% | RDT-based MSAT and fMDAs, snowball, pooled PCR |
| Bbondo, 9.2% | MSAT, household fMDA | MDA, snowball, all other fMDAs | MDA, serology-based household fMDA if coverage > 90%, serology-based MSAT if coverage > 95% | Pooled PCR, RDT-based MSAT and fMDAs, snowball |
| Munyumbwe, 27% | Household fMDA | MDA, MSAT, snowball, all other fMDAs | Elimination not realistic with single round of drug campaign |  |
| Sinamalima, 49% | Household fMDA | MDA, MSAT, snowball, all other fMDAs | Elimination not realistic with single round of drug campaign |  |

The spatial clustering of malaria infections means that fMDA strategies outperform MDA at selective targeting of infected individuals. Shared household exposure can arise from both features of geography—local availability of larval habitat—and of human behavior—household preference for insecticide-treated net (ITN) use and shared travel history. Due to the absence of data on individual histories of ITN use and travel, we assumed all infections were due to locally-acquired infections, and ITN usage was implicitly accounted for in each household's selected transmission intensity.

How crucial is avoiding overtreatment with antimalarial drugs? MDA is the most effective infection-detection strategy for both control and elimination, yet MDA also leads to the most overtreatment. When a drug campaign is a last push toward elimination and unlikely to be repeated many times, overtreatment may be less of an issue, particularly if the campaign is set up for success with high coverage and a long-lasting prophylactic. Given an excellent prophylactic, overtreatment is an irrelevant concern for elimination, particularly if vectors can migrate considerable distances. If the drug campaign is for control purposes, for instance as a stopgap measure when health systems are temporarily broken as during the recent Ebola outbreak [33], or as an ongoing program for gradual reduction in burden, minimizing overtreatment should be more of a priority. Our recommendations for optimal infection detection strategies prioritize avoiding overtreatment for control recommendations and use overtreatment as a secondary consideration for elimination campaigns.

Local prevalence, household density, and heterogeneity of RDT positivity all influence the optimal infection-detection strategy. While prevalence and population density may be known or estimated prior to a campaign, describing regional heterogeneity in exposure often requires more investment of resources through ongoing longitudinal surveys, multi-antigen serology, or sequencing of parasite genomes [15,34-36]. Local population density and entomology can guide planners' choice of fMDA radius if fMDA is under consideration.

In all but the lowest-prevalence settings, coverage is a stronger determinant of campaign success than the choice of infection-detection strategy. However, when transmission is very low, the limited sensitivity of current diagnostics means that index cases are unlikely to be discovered by current RDTs, and MDA, highly sensitive techniques like pooled PCR, or serological diagnostics that integrate history of infection are required to significantly reduce onward transmission. Greater coverage cannot compensate for an insensitive diagnostic. Simulation of serology-based diagnostics suggest that it is indeed possible to interrupt transmission in low-prevalence regions without distributing prophylactics to all individuals in the elimination area, although this finding may vary widely according to local entomology.

Under moderate to high prevalence, achieving high coverage is more important than selecting the optimal campaign type. Of the non-MSAT strategies, all are equally efficacious at high prevalence, and within-household fMDA results in the least overtreatment. When transmission is moderate, both MSAT and within-household fMDA are viable options, and other considerations such as cost, feasibility, and local culture will play a larger role in identifying the optimal infection detections strategy.

Compared with fMDAs, snowball RCD is a poor infection-detection strategy at moderate prevalence. In snowball RCD, an initial clinical case serves as the primary trigger for a 200m-fMDA, and each RDT positive in that 200m radius triggers a secondary rounds of 200m-fMDA. In low-transmission settings (Gwembe HFCA), new infection is likely to lead to a clinical case, and thus a primary trigger, and although secondary triggers are uncommon due to low prevalence and old infections, there are enough primary triggers to achieve good spatial coverage in local areas of transmission. In high-transmission settings (Sinamalima HFCA), new infection is unlikely to lead to a primary trigger, but secondary triggers are common and thus the snowball effect leads to good spatial coverage and behaves like fMDA. Under moderate transmission, infections are unlikely to lead to primary triggers due to immunity to clinical symptoms, and secondary triggers are less common than in high-transmission settings because infections are older and less likely to be superinfections. In addition, symptomatic individuals are often less likely to seek care if they live further from a clinic [37], leading to spatial dependence in detecting primary triggers.

In this study we have approximated interruption of transmission as the probability of less than one new infection per 1000 people arising from untreated infections in the 30 days post-campaign. In a more realistic scenario, multiple campaign rounds per year are carried out, and campaigns may last for several years. Thus we expect that all infection detection strategies are potentially more effective for elimination than predicted in the single-round analysis, but their relative efficacy will be as described above. Other modeling studies have suggested that multiple rounds of drug campaigns in moderate-prevalence settings such as Munyumbwe HFCA may successfully interrupt transmission [5]. In addition, programs are also likely to adapt campaign strategies as more data is collected, local pockets of transmission are identified, and overall prevalence declines. A full dynamical model of malaria transmission at the household scale, with detailed simulation of vector feeding behavior and movement at the individual vector level, is necessary to fully explore the elimination power of various infection detection strategies.

Selective targeting of hotspots of malaria transmission has been proposed as a control measure, yet correctly identifying hotspots remains a challenge with current tools, particularly in seasonal settings where drug campaigns are likely to be deployed during the low transmission season. We predict that selective targeting via MSAT or fMDA strategies will not succeed in elimination until a new generation of diagnostics is ready for field use. While our study does not rule out the possibility that repeated targeting of hotspots over many years may eventually lead to elimination, such extended campaigns pose significant feasibility challenges to communities, programs, and donors.

Size of the local population and patterns of human migration affect the likelihood of elimination, as elimination is easier in smaller population pools with less human mobility. Human mobility and spatial heterogeneity also interact to inform local prevalence in complex ways [38]. A study of RCD in a low-transmission region of Senegal found that most index cases reported recent travel [27]. If every RDT positive case identified in the Gwembe HFCA were due to household inhabitants migrating from higher transmission regions, all RDT-dependent infection detection strategies would be less successful (Fig S7). People arriving from higher-transmission areas will have relatively stronger immune responses to infection, making those infections more difficult to detect. Infections detected in Gwembe HFCA could also be due to Gwembe inhabitants traveling elsewhere, acquiring infections, and returning to Gwembe; in this scenario, these individuals would be more likely to harbor recent, high-density infections amenable to detection by RDT. Understanding regional demographics of mobility and inter-connectedness of elimination candidate areas can lead planners to decide whether non-MDA infection detection strategies are viable alternatives, and whether MSAT or MDA at border crossings would be effective policies [39].

This study ignores campaign cost and feasibility as considerations for selecting an infection-detection strategy, yet these factors are important drivers in the real world. Operational limitations make MDA and MSAT more attractive options than fMDAs, RCD, and strategies that require sensitive but expensive diagnostics. In elimination scenarios, achievability may overrule cost as a consideration for determining campaign strategy. Fortunately we find that MDA and MSAT are already the best strategies for elimination and control respectively in low-prevalence settings where drug campaigns are most likely to be deployed.

## Materials and Methods

### Estimation of household transmission intensities

Reference data for household location, age structure, and RDT positivity by age was derived from a 2012 June-July survey performed in Gwembe and Sinazongwe districts in Southern Province, Zambia [19]. Malaria transmission is heterogeneous and seasonal, with peak transmission between March and May. Households in four HFCAs—Gwembe, Bbondo, and Munyumbwe in Gwembe district as well as Sinamalima in Sinazongwe district—were selected for inclusion in the reference dataset (Fig 1,

Table 1, Fig S1, Dataset S1). Households without geolocation data and individuals without an RDT result were excluded.

Each household's exposure to infectious bites was determined as follows. An agent-based mechanistic model of malaria transmission, including exposure-dependent host immunity, was used to generate simulated populations experiencing endemic transmission (EMOD DTK v2.0) [40-43]. Twelve simulations of 10,000 people were run, where each simulation experienced the same southern Zambia seasonal temperature and rainfall patterns but supported different amounts of vectors (Fig S2). These twelve simulations spanned a range of annual entomological inoculation rates (EIRs) from 0.003 to 120 infectious bites per person per year and included 10 imported cases per year. All simulations incorporated case management as 30% treatment rate of clinical malaria and 50% treatment rate of severe malaria with artemether-lumefantrine. Vector control was implicitly modeled in household exposure, and within-household heterogeneity in use of ITNs was ignored. Simulations recorded daily asexual parasite density, infectiousness, and fever temperature for all individuals. Infectiousness was defined as the fraction of mosquitoes feeding on the individual that would become infected and develop at least one oocyst. Asexual parasite density and infectiousness were previously calibrated to age-stratified data from Burkina Faso [22]. Each simulation was run for 90 years, allowing births and deaths but holding total population fixed, and RDT prevalence by age was measured on June 15 of year 90, with RDT sensitivity at 50 asexual parasites/μL. A higher RDT sensitivity was used here compared to later simulations as community health workers who gathered the reference dataset were highly trained in RDT use.

The relative probability $P_{ij}$ that a household $i$ experiences exposure modeled by simulation $j$ was calculated as follows for each household in the reference dataset and each of the twelve reference simulations. All individuals $k$ within 50m of the household were assumed to experience similar transmission intensity and aggregated to better inform household exposure. The fraction of people of $k$'s age $a_k$ in simulation $j$ with $k$'s RDT positivity, $R_j^{\pm}(a_k)$, is multiplied over all $k$ within 50m of the household to find $P_{ij}$:

$$P_{ij} = \prod_{\substack{\text{individuals } k}}^{\text{within 50m of household } i} R_j^{\pm}(a_k) \tag{1}$$

Household transmission intensity is then determined by random selection from the $j$ simulations according to weights $P_{ij}$.

After selecting household transmission intensity, individuals were drawn from the 10,000 individuals simulated at that transmission intensity to form the age and RDT-positive structure of the household observed in the reference dataset. For example, if an RDT-positive 6-year-old was observed in the dataset household, an RDT-positive 6-year-old was drawn from the simulation pool, and so on until the household was complete. Household construction and infection detection campaigns were carried out 100 times per coverage level per infection-detection strategy for each HFCA to allow for stochastic variation in selecting transmission intensity, selecting individuals from simulations, and individual coverage during the drug campaign.

**Simulation of drug campaigns**

The following infection detection strategies were tested in the four HFCAs: MDA, MSAT, within-household fMDA—treating all individuals in the same household as someone testing positive, within-50m fMDA—treating all individuals within 50m of someone testing positive, within 200m-fMDA—treating all individuals within 200m of someone testing positive, and snowball RCD—treating all individuals within 200m of individuals with temperature > 38.5°C, testing all individuals within 200m of individuals with fever > 1.5°C, and treating all individuals within 200m of someone testing positive. Pooled PCR, where blood spots from neighboring households are pooled prior to a PCR-based diagnosis [44], and serological MSAT and fMDA were also tested in the low-prevalence HFCAs of Gwembe and Bbondo. For pooled PCR, households were divided into pools by eye according to spatial proximity such that each pool contained 60-220 individuals (Fig S3). See Table 2 for definitions of infection detection strategies tested.

All strategies other than snowball RCD were simulated as being carried out on a single day, June 15. Snowball RCD was carried out daily for 30 days, June 15-July 15, with completely correlated coverage: for example, under 70% coverage, the same 30% of individuals are unreachable every day of the campaign. All strategies were tested with coverage 0-100% at 5% intervals. Coverage was determined by individual rather than by household and was age- and location-independent.

The drug used for campaigns was a hypothetical drug that targeted both asexual and sexual stages and provided prophylactic protection for more than one month, similar to the combination therapy dihydroartemisinin-piperaquine. All covered individuals were assumed to accept and fully comply with treatment, and treatment was assumed to clear all asexual and sexual stage parasites in one day. RDT sensitivity was assumed to be 100 asexual parasites/µL during the test campaigns [45].

**Estimation of onward transmission**

To calculate the effects of drug campaigns on onward infection, the expected number of new infections in the human population that would arise from vectors infected in the 30 day period June 15-July 15 was estimated as follows:

$$\text{new infections per } 1000 = \frac{1000}{\text{population}} \left( \sum_{\text{population}} \sum_{30 \text{ days}} I * HBR * F \right) * P \qquad (2)$$

where *I* is the daily individual infectiousness, *HBR* is the daily individual human biting rate and differs according to household transmission intensity and individual age, and *P* is the product of the probability the mosquito survives feeding (0.9) [40], probability oocysts survive into sporozoites (0.8) [22,46], probability a mosquito survives sporogony at 20°C (0.15) [40,47], and probability an infectious mosquito bite successfully infects a human (0.9) [48]. Individual infectiousness is 0 upon treatment. Treated individuals were assumed to be protected by prophylaxis, and thus only bites on untreated individuals could lead to new infections. *F* is the probability that the recipient of an infectious bite is susceptible to infection:

$$F = \frac{\sum_{\text{untreated}} \sum_{30 \text{ days}} HBR e^{-x^2/0.2^2}}{\sum_{\text{population}} \sum_{30 \text{ days}} HBR\, e^{-x^2/0.2^2}} \tag{3}$$

*F* was defined as the fraction of bites that occur on untreated individuals, where each individual's probability of receiving a bite is weighted by their distance *x* from the individual who infected the mosquito. The mosquito is assumed to diffuse an expected distance of 0.2 km within the 30 days of biting. The new infection rate was normalized to a population of 1000 to facilitate comparison across HFCAs.

To predict the number of clinical cases arising from the estimated number of new infections, birth cohort simulations were run for a duration of twenty years across a wide range of transmission intensities (annual EIR from 0.01 to 128) and with twelve different initialization dates, each corresponding to the first day of each calendar month. New infections and new clinical cases for all days in the range June 15-July 15 were tallied separately by age and by EIR. Multipliers were then computed from the ratio of the distribution means to derive age- and EIR-specific transformations of new infections to clinical cases.

For estimating the probability a campaign leads to less than one new infection, coverage was tested from 50-100% at 1% intervals. Each infection-detection strategy and coverage level was simulated for 1000 stochastic realizations, and the number of simulations where Equation 2 evaluated to < 1 was counted.

**Acknowledgments**

We would like to acknowledge the support of the Zambia Ministry of Health, National Malaria Control Centre, the Provincial Health Office, and local district health offices in conducting the data collection. We are also grateful to the communities involved in the research for the continued support in the quest to eliminate malaria.

**Supporting Information**

**Figure S1.** Demographic and geographic features of study area households.

**Figure S2.** Simulation RDT prevalence by age and seasonal malaria transmission.

**Figure S3.** Neighborhood pools used for pooled PCR testing.

**Figure S4.** Campaigns without a long-lasting prophylactic are much less successful at averting new infections.

**Figure S5.** Outcomes of RDT-dependent campaigns in Gwembe HFCA using an improved RDT with sensitivity of 10 parasites/μL.

**Figure S6.** Likelihood of interrupting transmission increases if vectors are not assumed to remain close to their source of infection.

**Figure S7.** Infection detection in Gwembe HFCA if all infections are importations.

**Dataset S1.** Individual RDT positivity by household, age, location, and HFCA.

Figure 1

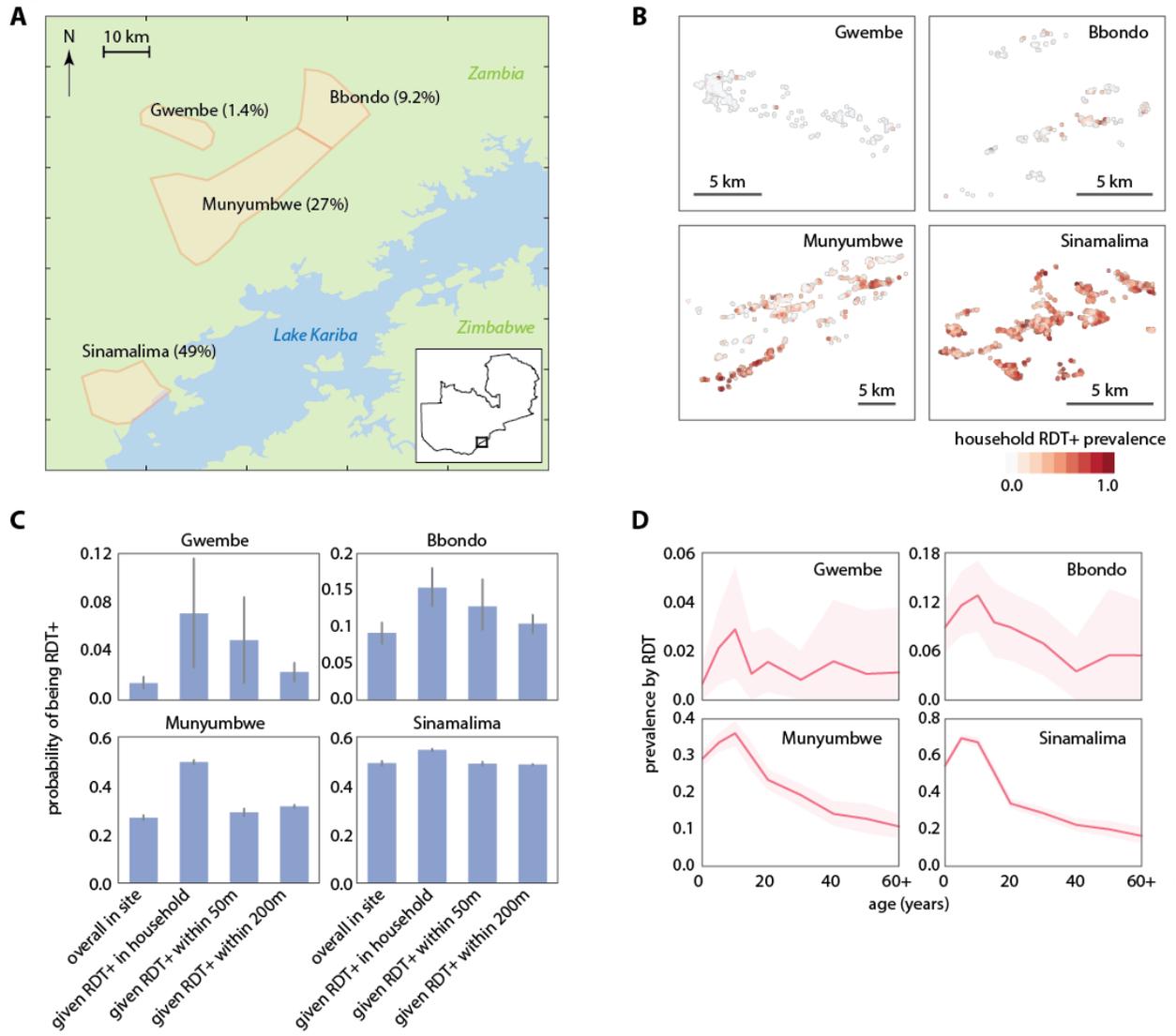

Figure 2

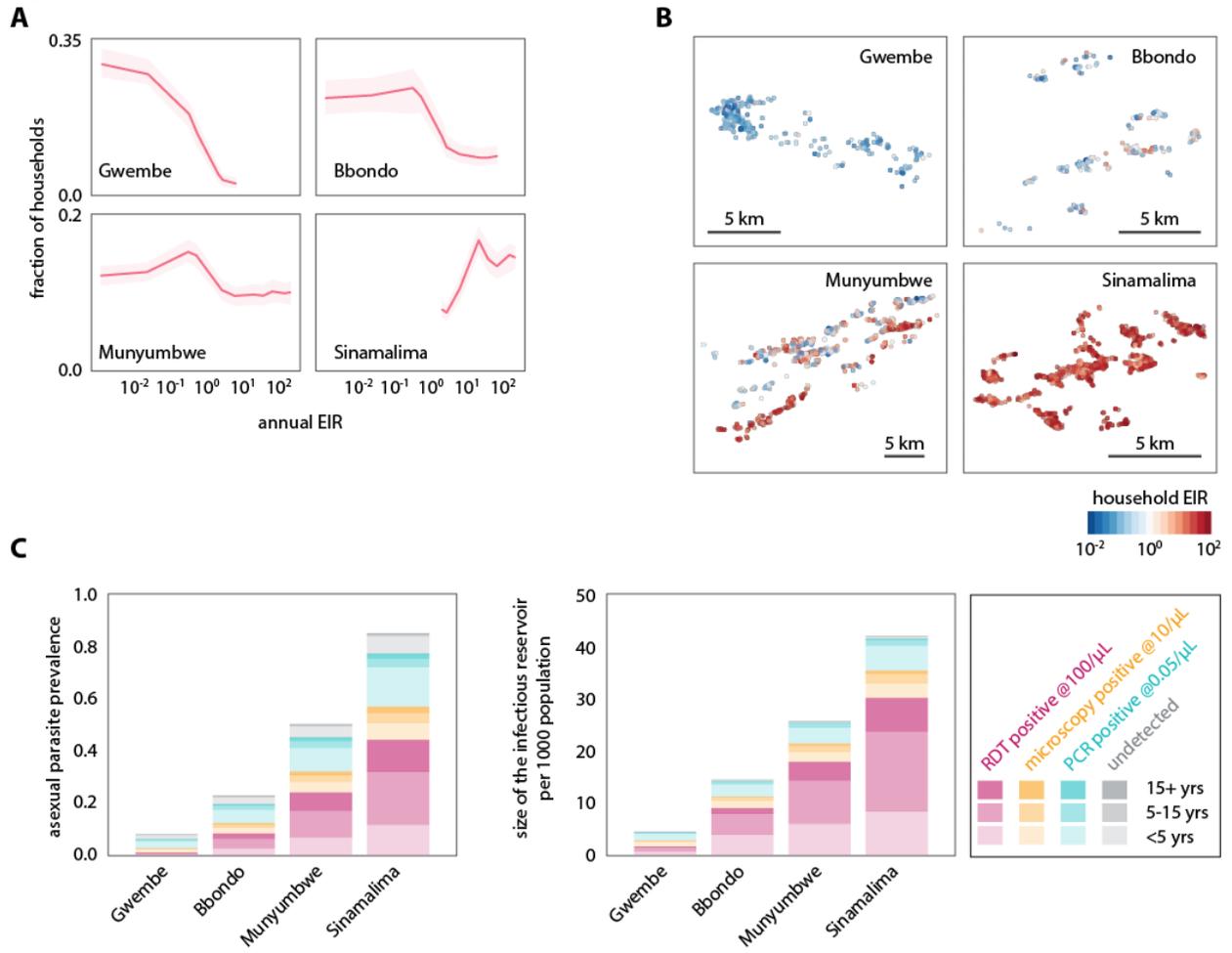

Figure 3

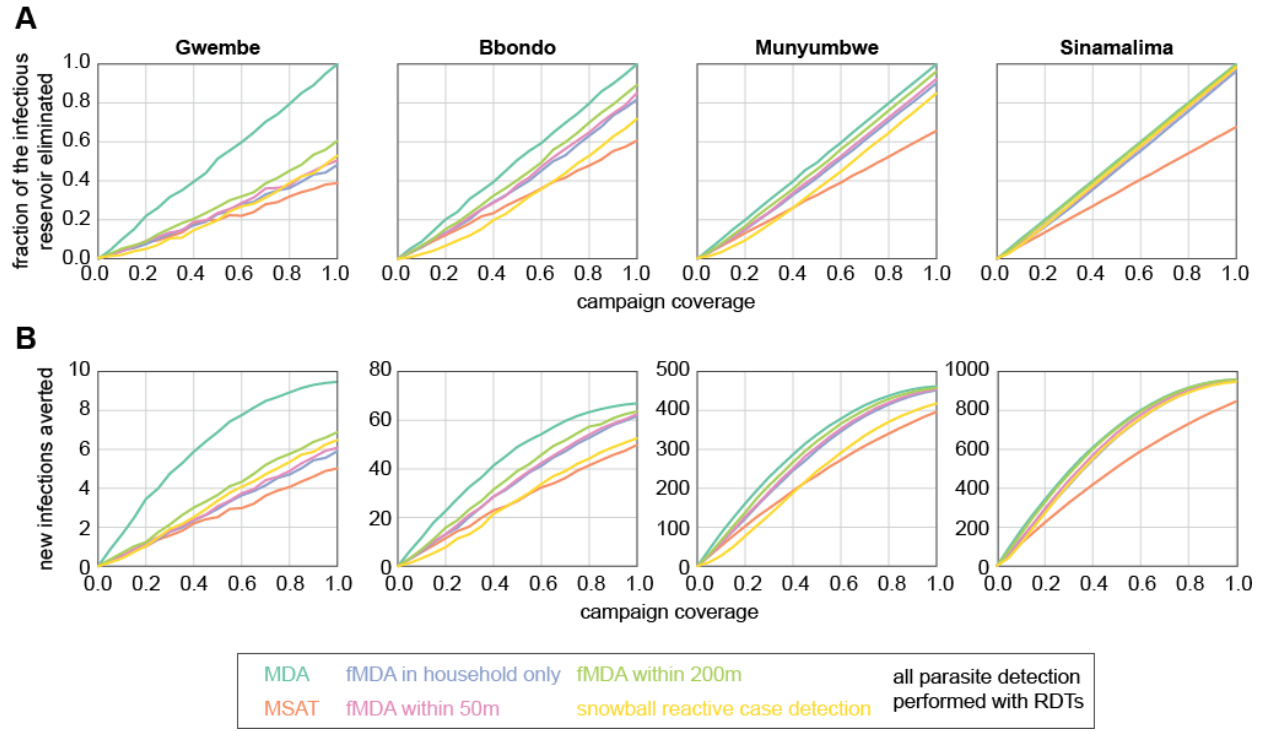

Figure 4

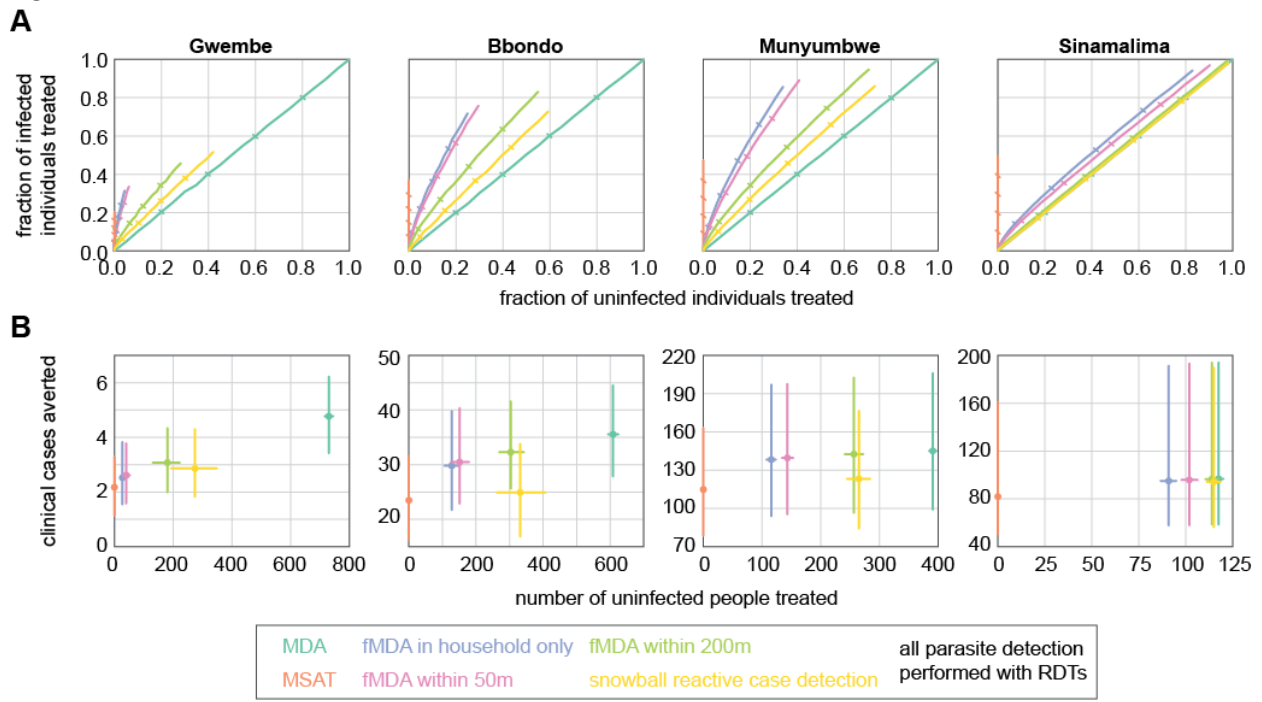

Figure 5

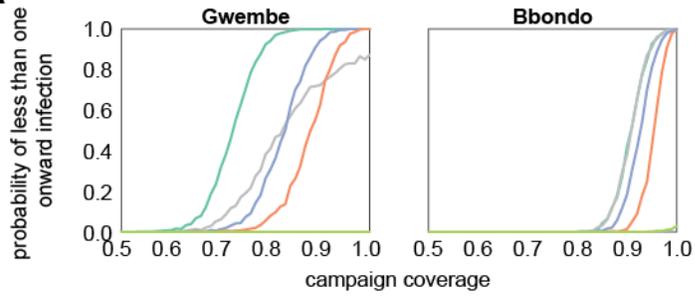

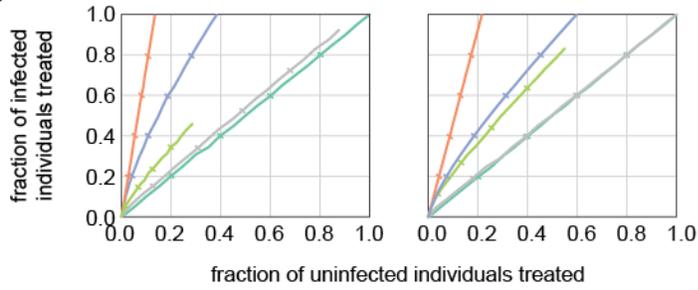

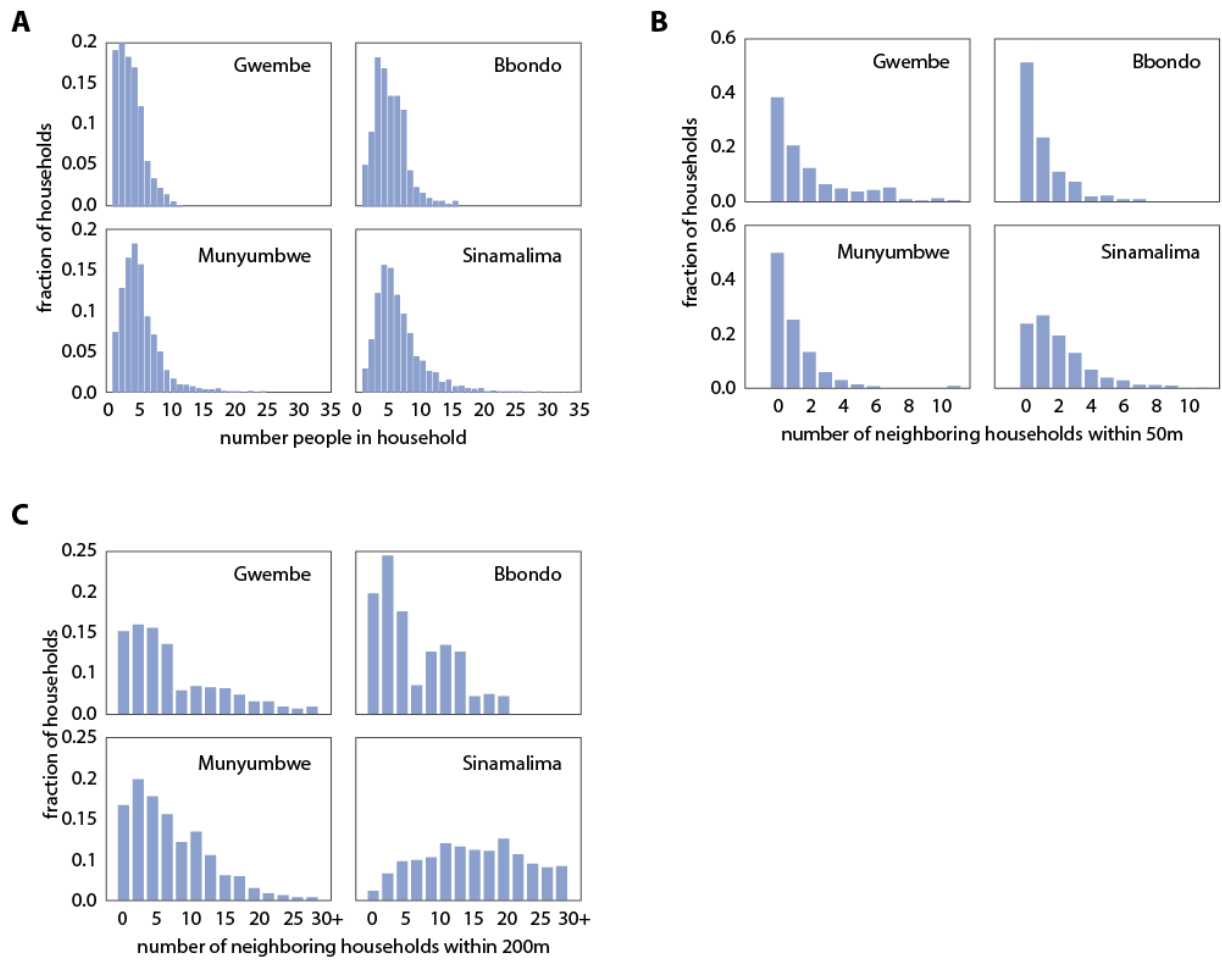

Figure S1. Demographic and geographic features of study area households. (A) Distribution of number of people per household by HFCA. (B) Distribution of number of neighboring households within 50m of each household by HFCA. (C) Distribution of number of neighboring households within 200m of each household by HFCA.

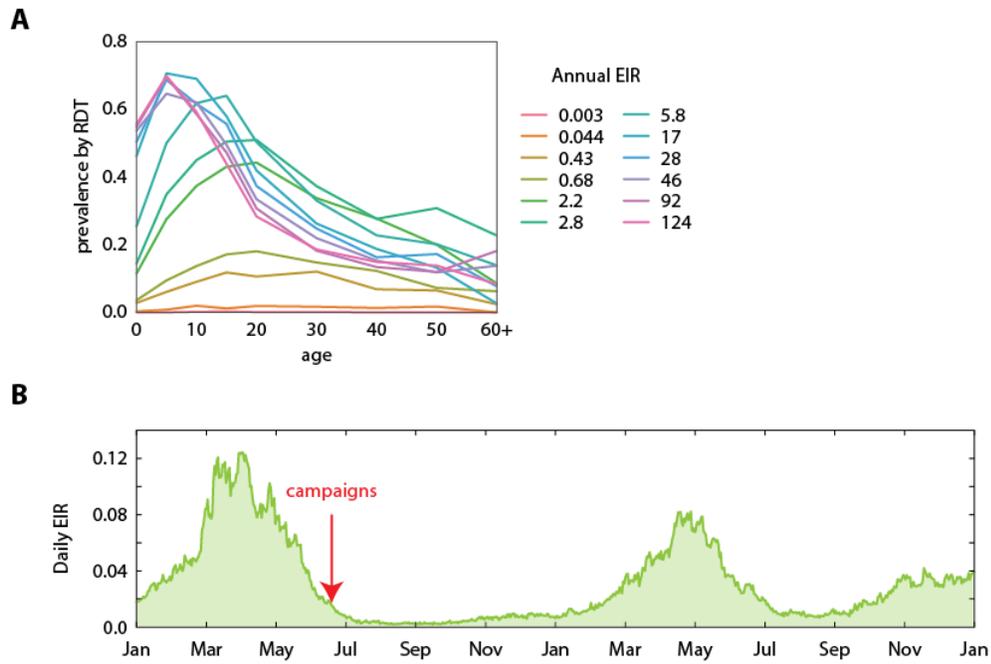

Figure S2. Simulation RDT prevalence by age and seasonal malaria transmission. (A) RDT prevalence by age on June 15 for twelve simulated transmission intensities. (B) Seasonal malaria transmission modeled under southern Zambia climate. Drug campaigns take place on June 15, at the beginning of the dry season.

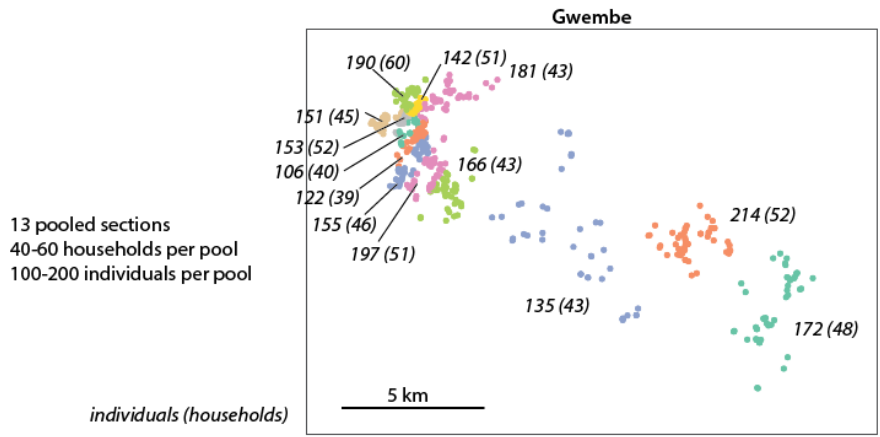

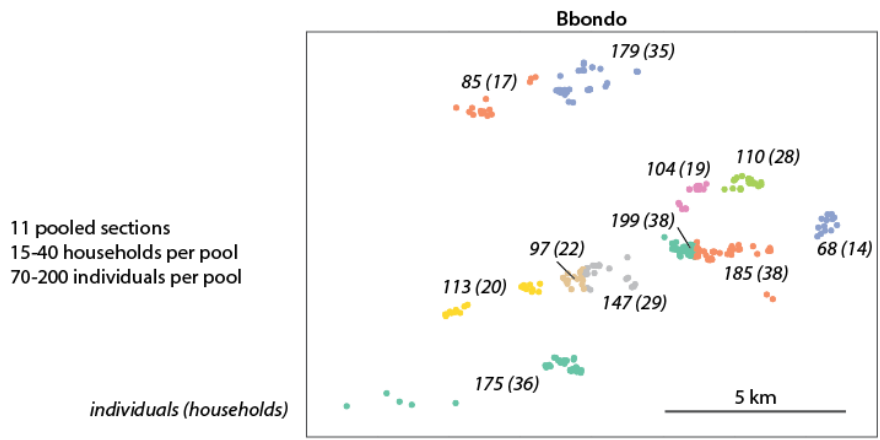

Figure S3. Neighborhood pools used for pooled PCR testing in (A) Gwembe HFCA and (B) Bbondo HFCA.

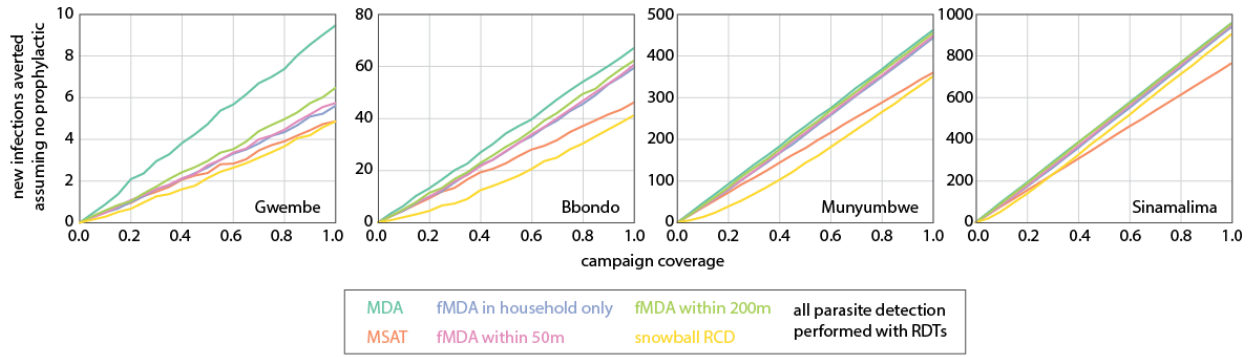
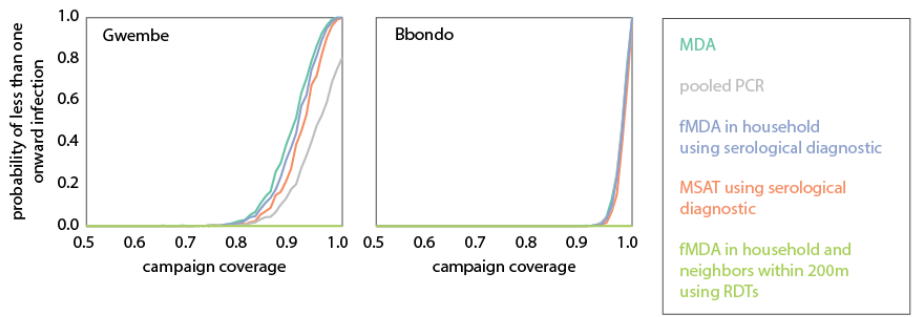

Figure S4. Campaigns without a long-lasting prophylactic are much less successful at averting new infections. (A) Success of infection detection strategies at averting new infections when campaign drug is not prophylactic. Mean of 100 stochastic realizations per coverage level. HFCA populations normalized to 1000. (B) Probability of fewer than 1 onward infection per 1000 people in Gwembe and Bbondo HFCAs if the campaign drug has no prophylactic effect. Mean of 1000 stochastic realizations.

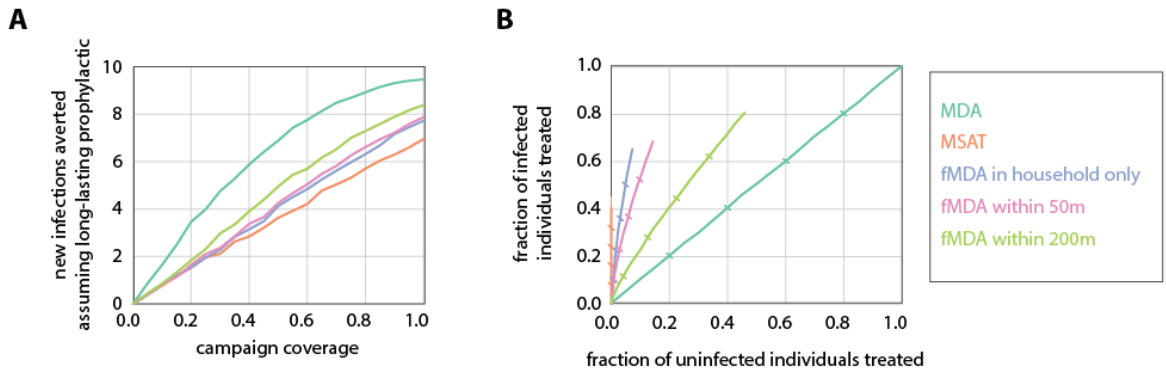

Figure S5. Outcomes of RDT-dependent campaigns in Gwembe HFCA using an improved RDT with sensitivity of 10 parasites/μL. (A) Success of infection detection strategies at averting new infections. Mean of 100 stochastic realizations per coverage level. HFCA populations normalized to 1000. (B) Success of infection detection strategies at finding infected individuals while minimizing overtreatment. Mean of 100 stochastic realizations per coverage level. Ticks indicate every 20% of coverage.

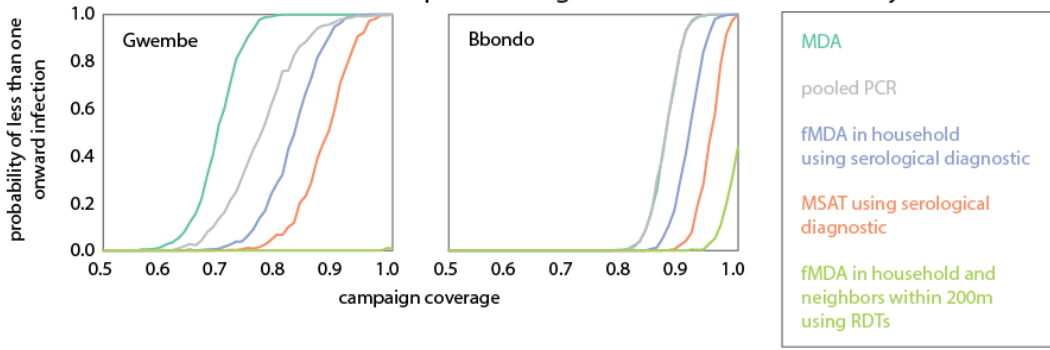

Figure S6. Lkelihood of interrupting transmission increases if vectors are not assumed to remain close to the source of their infection. Probability of fewer than 1 onward infection per 1000 people in Gwembe and Bbondo HFCAs if infected vectors are free to bite anyone in the HFCA rather than tending to remain close to the household where they were infected (all weights set to 1 in equation 3). Mean of 1000 stochastic realizations.

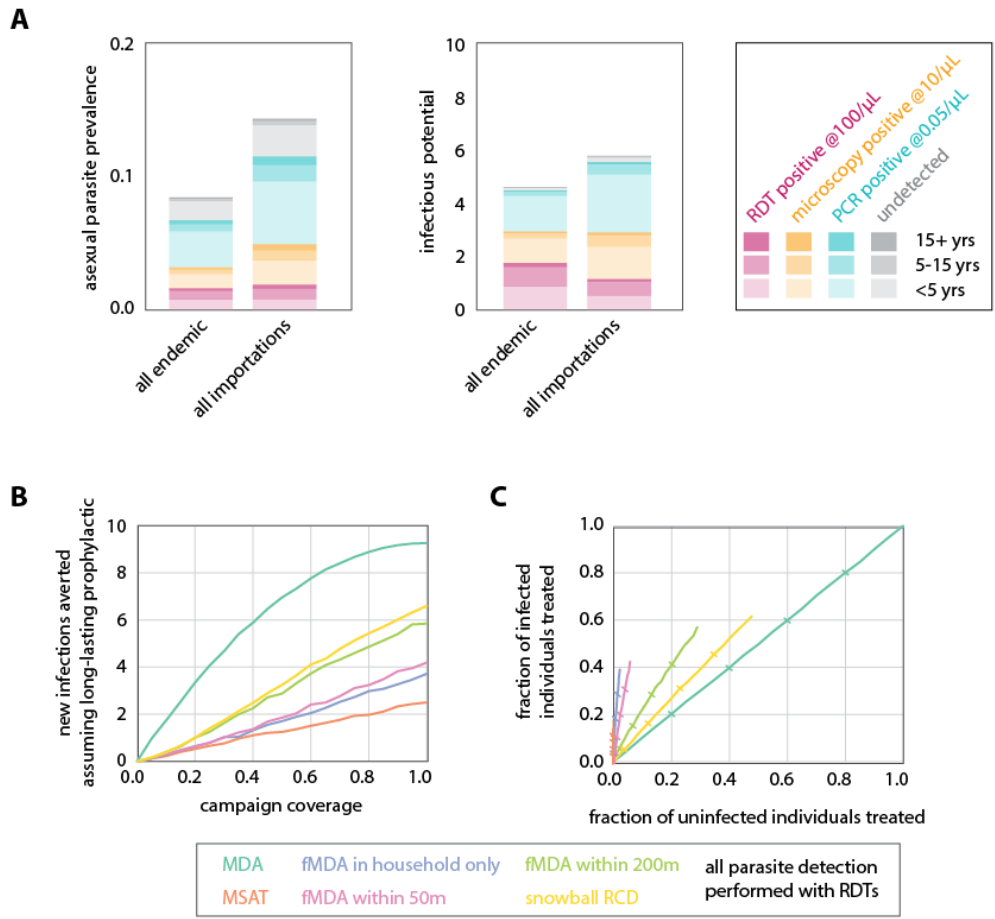

Figure S7. Infection detection in Gwembe HFCA if all infections are importations. For the imported case, all members of households with at least one RDT+ individual are modeled as recent immigrants from a region of EIR = 50. (A) True asexual parasite prevalence and infectious potential if all RDT positive infections are endemic or imported. Infectious potential defined as in Figure 2C. Mean of 100 stochastic realizations. (B) Success of infection detection strategies at averting new infections. Mean of 100 stochastic realizations per coverage level. HFCA population normalized to 1000. (C) Success of infection detection strategies at finding infected individuals while minimizing overtreatment. Mean of 100 stochastic realizations per coverage level. Ticks indicate every 20% of coverage.